\documentclass[11pt]{article}
\newcommand{\Comment}[1]{{}}
\usepackage{amsfonts,amsmath,epsfig,color,latexsym,graphics}
\usepackage[textwidth = 430 pt, textheight = 630 pt]{geometry}

\Comment{\usepackage{color}
\definecolor{MyDarkBlue}{rgb}{0.15,0.15,0.45}
\usepackage[linktocpage=true]{hyperref}
\hypersetup{
colorlinks=true,
citecolor=MyDarkBlue,
linkcolor=MyDarkBlue,
urlcolor=MyDarkBlue,
pdfauthor={Horatiu Nastase and Howard J. Schnitzer},
pdftitle={Twistor and Polytope Interpretations for Subleading Color One-Loop Amplitudes},
pdfsubject={hep-th}
}

\usepackage[numbers,sort&compress]{natbib}
\usepackage{hypernat}}

\newcommand\ignore[1]{}
\def\one{{\,\hbox{1\kern-.8mm l}}}

\def\Tr{{\rm Tr\, }}

\def\a{\alpha}\def\b{\beta}

\def\C{\mathbb{C}}

\newcommand{\Cset}{{\,\,{{{^{_{\pmb{\mid}}}}\kern-.45em{\mathrm C}}}}}

\newcommand{\be}{\begin{equation}}
\newcommand{\bea}{\begin{eqnarray}}

\newcommand{\ee}{\end{equation}}
\newcommand{\eea}{\end{eqnarray}}

\begin{document}

\renewcommand{\thefootnote}{\fnsymbol{footnote}}

\makeatletter
\@addtoreset{equation}{section}
\makeatother
\renewcommand{\theequation}{\thesection.\arabic{equation}}

\rightline{}
\rightline{}
   \vspace{1.8truecm}

\begin{flushright}
BRX-TH-634
\end{flushright}

\vspace{10pt}


\begin{center}
{\LARGE \bf{\sc Twistor and Polytope Interpretations \\for Subleading Color One-Loop Amplitudes
}}
\end{center} 
 \vspace{1truecm}
\thispagestyle{empty} \centerline{
    {\large \bf {\sc Horatiu Nastase${}^{a,}$}}\footnote{E-mail address: \Comment{\href{mailto:nastase@ift.unesp.br}}{\tt 
    nastase@ift.unesp.br}} {\bf{\sc and}}
    {\large \bf {\sc Howard J. Schnitzer${}^{b,}$}}\footnote{E-mail address:
                                \Comment{ \href{mailto:schnitzr@brandeis.edu}}{\tt schnitzr@brandeis.edu}}
                                                           }

\vspace{1cm}
\centerline{{\it ${}^a$ 
Instituto de F\'{i}sica Te\'{o}rica, UNESP-Universidade Estadual Paulista}} \centerline{{\it 
R. Dr. Bento T. Ferraz 271, Bl. II, Sao Paulo 01140-070, SP, Brazil}}

\vspace{.8cm}
\centerline{{\it ${}^b$ 
Theoretical Physics Group, Martin Fisher School of Physics}} \centerline{{\it Brandeis University, Waltham, 
MA 02454, USA}}

\vspace{2truecm}

\thispagestyle{empty}

\centerline{\sc Abstract}

\vspace{.4truecm}

\begin{center}
\begin{minipage}[c]{380pt}{\noindent We use the relation of the one-loop subleading-color amplitudes to the one-loop $n$-point leading color
amplitudes in ${\cal N}=4$ SYM, to derive a polytope interpretation for the former in the $MHV$ case, and a representation in momentum 
twistor space for the general $N^kMHV$ case. These techniques are explored in detail for the 5-point and 6-point amplitudes.
We briefly discuss the implications for IR divergences. 
}
\end{minipage}
\end{center}

\vspace{.5cm}

\setcounter{page}{0}
\setcounter{tocdepth}{2}

\newpage

\renewcommand{\thefootnote}{\arabic{footnote}}
\setcounter{footnote}{0}

\linespread{1.1}
\parskip 4pt

{}~
{}~

\section{Introduction}

There has been significant progress in the understanding and efficiency of calculation of perturbative scattering amplitudes for 
${\cal N}=4$ SYM theory in the planar limit. Important advances have made use of twistors, momentum twistor space and dual conformal symmetry 
(DCI) to provide an understanding of the integrand of the $n$-point, multiloop amplitudes, e.g. 
\cite{ArkaniHamed:2010gg,Mason:2010pg,ArkaniHamed:2010gh,ArkaniHamed:2009dn,ArkaniHamed:2009vw,Mason:2009qx,Hodges:2009hk,Alday:2009zm,Hodges:2010kq,Britto:2004nc,ArkaniHamed:2010kv,ArkaniHamed:2009sx,ArkaniHamed:2009si,Drummond:2009fd,Drummond:2008vq,Drummond:2008cr,Drummond:2010qh,Drummond:2010uq,Drummond:2008bq,Drummond:2010mb,Drummond:2006rz,Hodges:2005bf,Hodges:2005aj,Hodges:2006tw,Bullimore:2010pj,Boels:2007qn,Mason:2010yk,Mason:2009sa,Bullimore:2010dz,Bullimore:2010pa,Korchemsky:2010ut,Korchemsky:2009jv,Beisert:2010gn,He:2010ju,Kaplan:2009mh,Boels:2010nw,Vergu:2009tu,Bern:1994zx,Brandhuber:2008pf,Brandhuber:2007yx,Brandhuber:2010mi,Brandhuber:2005kd,Brandhuber:2004yw}. 
These advances have given rise to elegant 
geometric interpretations of integrands in terms of polytopes obtained from gluing of elementary simplices \cite{ArkaniHamed:2010gg}. For MHV $n$-point 
planar amplitudes at one-loop, one can also give a geometric interpretation of the integrated amplitudes \cite{Mason:2010pg}. Here again a 
polytope picture emerges, with the IR divergences controlled by a mass regulator. These various geometric descriptions give new insights into the 
structure of the theory. 

In this paper we make use of twistor methods and polytope interpretations of subleading one-loop color-ordered amplitudes for $MHV$, $NMHV$ and more 
generally $N^kMHV$ amplitudes. Although one-loop subleading-color amplitudes are determined by the leading color amplitudes, there is significant 
motivation to consider these subleading color amplitudes in their own right. Firstly, rich polytope structures emerge from the analyses which 
involve gluings of the polytopes of the leading color amplitudes. As a corollary of the analysis, we show that the leading IR divergence of the 
one-loop subleading amplitudes are ${\cal O}(1/\epsilon)$ in contrast to the $1/\epsilon^2$ IR divergence of the planar amplitudes. 

Interesting connections have been made between subleading ${\cal N}=4$ SYM amplitudes and these of ${\cal N}=8$ sugra at the one- and two-loop level. 
Connections between the two theories are very intriguing, and a deeper understanding of subleading amplitudes in ${\cal N}=4$ SYM may contribute 
insights into the structure of ${\cal N}=8$ sugra. This could be useful, as the technology for perturbative calculations for ${\cal N}=8$ sugra lags 
behind that of ${\cal N}=4$ SYM and other gauge theories. 

The improvement in calculational technologies for ${\cal N}=4$ SYM is now making an impact on QCD calculations \cite{Berger:2010zx,Dixon:2010ik}. 
With that in mind, the $SU(3)$ gauge 
group of QCD may require $1/N$ corrections to the planar limit in applications. Therefore advances in our understanding of the non-planar corrections
to ${\cal N}=4$ SYM could be useful in QCD applications as well. These all point to the importance of understanding subleading amplitudes.

The color-dressed tree amplitude ${\cal A}_n$ of ${\cal N}=4$ SYM is related to the color-ordered tree amplitude $A_n$ 
by (using the notation in \cite{Bern:2008qj})
\bea
{\cal A}_n^{tree}(12...n)&=&g^{n-2}\sum_{\sigma\in S_n/Z_n}\Tr(T^{a_{\sigma(1)}}...T^{a_{\sigma(n)}})A_n^{tree}(\sigma(1)...\sigma(n))\cr
&=&g^{n-2}\sum_{P(23...n)}\Tr[T^{a_1}T^{a_{P(2)}}...T^{a_{P(n)}}]A_n^{tree}(12...n)
\eea
where $1$ is fixed and $P(23..n)$ is a permutation of $2,3,...,n$.

For the $n$-point color-ordered tree amplitudes, there is a basis of $(n-2)!$ amplitudes out of the total $n!$, 
called Kleis-Kuijf basis, and we can 
find the others easily in terms of it \cite{Bern:2008qj}. It is based on the existence of the Kleis-Kuijf relations (KK), which are 
\be
A_n(1,\{\a\},n,\{\b\})=(-1)^{n_\b}\sum_{\{\sigma\}_i\in OP(\{\a\},\{\b^T\})}A_n(1,\{\sigma\}_i,n)\label{KK}
\ee
where $\sigma_i$ are ordered permutations, i.e. that keep the order of $\{\a\}$ and of $\{\b^T\}$ inside $\sigma_i$. 
Thus the KK basis is $A_n(1,{\cal P}(2,...,n-1),n)$ where ${\cal P}$ are arbitrary permutations. We will use the KK basis as a convenient
expansion basis even at 1-loop, in the cases where we can factorize tree amplitudes. 

We note that, by using cyclicity, reflection invariance,
\be
A_n(12...n)=(-1)^nA_n(n...21)
\ee
and the KK relations (\ref{KK}), we can reconstruct all the $n!$ amplitudes from the KK basis.

At 1-loop the subleading pieces of the amplitude in the $1/N$ expansion can be obtained from the leading piece from \cite{Bern:1998sv}
\bea
{\cal A}_n^{1-loop}(12...n)&=&g^n \sum_{j=1}^{[n/2]+1}\sum_{\sigma\in S_n/S_{n;j}}G_{2n;j}(\sigma)A_{n;j}(\sigma(1)...\sigma(n))\cr
G_{rn;1}(1)&=&N_c\Tr(T^{a_1}...T^{a_n})\cr
G_{rn;j}(1)&=&\Tr(T^{a_1}...T^{a_{j-1}})\Tr(T^{a_j}...T^{a_n})\cr
A_{n;j}(12...,j-1,j,j+1,...n)&=&(-1)^{j-1}\sum_{\sigma\in COP\{\a\},\{\b\}}A_{n;1}(\sigma)\label{nonplanar}
\eea
where $S_{n;j}$ is the subset of permutations $S_n$ that leaves the trace structure $G_{rn;j}$ invariant, 
$\{\a\}=\{j-1,j-2,...,2,1\}$, $\{\b\}=\{j,j+1,...,n-1,n\}$ and $COP\{\a\},\{\b\}$ are permutations with $n$ fixed and keeping 
$\{\a\}$ and $\{\b\}$ fixed up to cyclic permutations. 

We will use these relations to find a polytope picture and a representation in momentum twistor space for the one-loop color 
ordered subleading amplitudes.

The paper is organized as follows. In section 2 we build a polytope picture for the $MHV$ subleading amplitudes. After reviewing the leading-order 
results, we first obtain the 5-point and 6-point subleading pieces, and then generalize to $A_{n;3}$ and $A_{n;j}$. In section 3 we build a 
twistor interpretation for the subleading pieces, after reviewing the leading order results. In section 4 we write explicit formulas for the 
6-point $NMHV$ amplitude, after writing a polytope picture for the leading order, and in section 5 we conclude. In the Appendices we write some 
long explicit formulas at 5-point and 6-point $MHV$ and analyze the IR divergences of the 6-point $NMHV$ amplitudes in $\mu$ regularization.

\section{Polytope methods for subleading amplitudes}

\subsection{Review of the method for leading MHV amplitudes}

In \cite{Mason:2010pg}, a simple picture was found for the 1-loop color ordered leading amplitudes of ${\cal N}=4$ SYM theory, 
in terms of the volume of a closed polytope in 
$AdS_5$. The starting point is writing amplitudes in a space dual to momenta, which trivializes the momentum conservation constraint, 
$\sum_i p_i=0$, by $p_i=x_i-x_{i+1}$. Then for instance the 1-loop dimensionless massless box function 
becomes (writing the loop momentum as $l=x_0-x_1$)
\be
F_{0m}(1,2,3,4)=i\int \frac{d^4x_0}{2\pi^2} \frac{(x_1-x_3)^2(x_2-x_4)^2}{(x_0-x_1)^2(x_0-x_2)^2(x_0-x_3)^2(x_0-x_4)^2}\label{box}
\ee
This representation has manifest conformal invariance in the dual space, or {\em dual conformal invariance} (DCI) \cite{Drummond:2010km}, 
although (\ref{box}) needs an infrared (IR) regulator. 
Following \cite{Mason:2010pg}, we then construct $x_{\a\dot\a}=x^\mu (\sigma_\mu)_{\a\dot\a}$ and finally map 
\be
x^{\a\dot\a}\rightarrow X^{AB}=\begin{pmatrix}-\frac{1}{2}\epsilon^{\a\b}x^2&ix^\a_{\dot\b}\\-ix^\b_{\dot\a}&\epsilon_{\dot\a\dot\b}\end{pmatrix}
\label{xtoX}
\ee
The $X$'s satisfy
\bea
&&X^2\equiv \frac{1}{2}\epsilon_{ABCD}X^{AB}X^{CD}=0\cr
&& X_i\cdot X_j =-(x_i-x_j)^2
\eea
and are coordinate patches on the quadric $X\cdot X=0$ in $RP^5$, with $X^{AB}\sim \lambda X^{AB}$ their homogeneous coordinates.
These $X$'s are considered as vertices situated at the boundary of an $AdS_5$ and are simple bitwistors living in twistor space, i.e. 
$\exists$ twistors $A^A$ and $B^B$ such that $X^{AB}=A^{[A}B^{B]}$ (a twistor $A^A$ is made of $(A^\a,A_{\dot\a})$). 

In the case of a box function characterized by $X_1,X_2,X_3,X_4$, the following  function of the Feynman parameters $\a_i\in (0,1)$ with $\sum \a_i=1$,
\be
X(\a)=\a_1X_1+\a_2X_2+\a_3X_3+\a_4X_4
\ee
is a map to $RP^5$, but such that $X(\a)\cdot X(\a)\neq 0$, and in fact they vary over a tetrahedron in $RP^5$.
After a normalization, 
\be
Y(\a)=\frac{X(\a)}{\sqrt{X(\a)\cdot X(\a)}}
\ee
one obtains $Y(\a)\cdot Y(\a)=1$, which means $Y(\a)$ lies in Euclidean $AdS_5$. Since straight lines $X(\a)$ are mapped to geodesics in $AdS_5$, 
the edges and faces of the tetrahedron in $AdS_5$ are geodesic, which by definition makes the tetrahedron {\em ideal}.

The value of the IR finite 4 mass box then matches twice the volume of the tetrahedron in $AdS_5$. The IR divergent lower mass functions 
need to be regularized, either in dimensional regularization, or a mass regularization as used in \cite{Mason:2010pg} that 
modifies $X^2=0$ to $X\cdot X= \mu^2 (X\cdot I)$, with $I$ a fixed point (A useful choice of $I$ is $X_i\cdot I=1, \forall i$). 

The one loop MHV $n-$point amplitudes divided by the tree MHV amplitudes are given by the sum of 1-mass and 2-mass easy box functions with 
coefficient one, which add up to the volume of a closed 3-dimensional polytope (without a boundary) with $n$ vertices. 

Note that here the definition of volume of a tetrahedron comes with a sign, determined by the order of the dual space vertices $x_i$ in  
the box function $F(i,j,k,l)$. That also induces an orientation (sign) for the triangular faces of the tetrahedron, determined by whether the 
missing vertex from $(ijkl)$ is in an even or odd position. Faces with same vertices and different orientation (sign) can be glued together, 
forming a continous object.

\subsection{Subleading 5-point and 6-point MHV amplitudes}

For the 5-point function, as explained in \cite{Mason:2010pg}, the leading 1-loop MHV amplitude $A_{5;1}^{MHV}(12345)$ divided by the tree amplitude
$A_5^{MHV}(12345)$ is the volume of the boundary of a 4-simplex,
\be
\frac{A_{5;1}^{MHV}(12345)}{A_5^{MHV}(12345)}\equiv M_5^{MHV}(12345)=\sum_{cyclic}I(x_1,x_2,x_3,x_4,(x_5))\equiv V(x_1,x_2,x_3,x_4,x_5)
\ee
Here $I(x_1,x_2,x_3,x_4,(x_5))$ is the volume of the tetrahedron with vertices $x_1,x_2,x_3,x_4$, equal to $F(1,2,3,4)$, and the missing vertex
$(x_5)$ is added 
in brackets since the cyclicity involves all 5 points; $V(x_1,x_2,x_3,x_4,x_5)$ is the volume of the boundary of the 4-simplex in $AdS_5$ space,
with $(y)_i\rightarrow (Y)_i$, i.e. we map the arguments of $V$ into $AdS_5$ space. One must be careful, since the $x\rightarrow X$ relation 
in (\ref{xtoX}) is not linear, so a linear combination  of $x_i$'s in not mapped to the linear combination of $X_i$'s. 

For $n=5$ there is only one subleading amplitude, $A_{5;3}$, which is related to the leading one by 
\be
A_{5;3}(45123)=\sum_{\sigma\in COP_4^{123}}A_{5;1}(\sigma(1),...,\sigma(4),5)
\ee
where $COP_4^{123}$ are cyclic permutations of $123$ inside 4 objects ($1234$).

The tree amplitudes can be expressed in terms of  the Kleis-Kuijf (KK) basis by the KK relations (\ref{KK}).
We can then express $A_{5;1}(12345)=A_5(12345)M_5(12345)$, etc. and write the tree amplitudes in the KK basis, obtaining for 
$A_{5;3}$
\bea
A_{5;3}(12345)&=&A_5(12345)[M_5(12345)-M_5(41235)+M_5(43125)-M_5(31245)]\cr
&&+A_5(12435)[M_5(12435)-M_5(31245)+M_5(34125)-M_5(41235)]\cr
&&+A_5(14235)[M_5(14235)-M_5(31425)+M_5(34125)-M_5(41235)]\cr
&&+A_5(13245)[M_5(23145)-M_5(31245)+M_5(43125)-M_5(42315)]\cr
&&+A_5(13425)[M_5(23145)-M_5(31425)+M_5(43125)-M_5(24315)]\cr
&&+A_5(14325)[M_5(23145)-M_5(31425)+M_5(34125)-M_5(23415)]\cr
&& \label{5pointKK}
\eea
We see 12 simplices appearing in the MHV case. Note that from $M_5^{MHV}(12345)=V(x_1,x_2,x_3,x_4,x_5)$ 
it does not follow that the permuted $M$'s have permuted $x$'s, 
since it is the momenta $p_i$ which are permuted, and that does not translate into permuted $x$'s. For example,\footnote{The paranthesis here 
do not mean missing vertices.} 
\be
M_5^{MHV}(23145)=V(x_1,(x_1-x_2+x_3),(x_1-x_2+x_4),x_4,x_5)
\ee
and we can check that the difference between consecutive arguments gives the permuted momenta, e.g. $p_2=x_1-(x_1-x_2+x_3)$. The full list 
of simplices is given in the Appendix.

Note that $V$ is invariant under a change of the origin of the $y_i$'s, so we can always add a constant to all the arguments. Using this 
freedom, we find for the coefficient of the KK basis element $A_5(12345)$ in (\ref{5pointKK})
\bea
&&M_5^{MHV}(12345)-M_5^{MHV}(41235)+M_5^{MHV}(43125)-M_5^{MHV}(31245)\cr
&=&V(x_1,x_2,x_3,x_4,x_5)-V((x_4-x_5+x_1),x_1,x_2,x_3,x_4)\cr
&&+V(x_4,(x_1+x_4-x_5),x_1,(x_1-x_3+x_4),(x_2-x_3+x_4))\cr
&&-V(x_1,(x_1-x_3+x_4),(x_2-x_3+x_4),x_4,x_5)\equiv V_1-V_2+V_3-V_4\label{termul}
\eea
We observe several facts. In this factor we have 
only 8 points: $x_1,x_2,x_3,x_4,x_5,x_6\equiv (x_1+x_4-x_5),x_8\equiv(x_2+x_4-x_3),x_7\equiv(x_1+x_4-x_3)$. 
There are two common points to all, $x_1$ and
$x_4$. One we fixed by translational invariance, but the second is nontrivial. More importantly, the two differences  have 4 common points out of 
5, so the differences in volumes are simpler. The full list of factors and their corresponding points is given in the Appendix.

\begin{figure}[h]
\begin{center}
\includegraphics{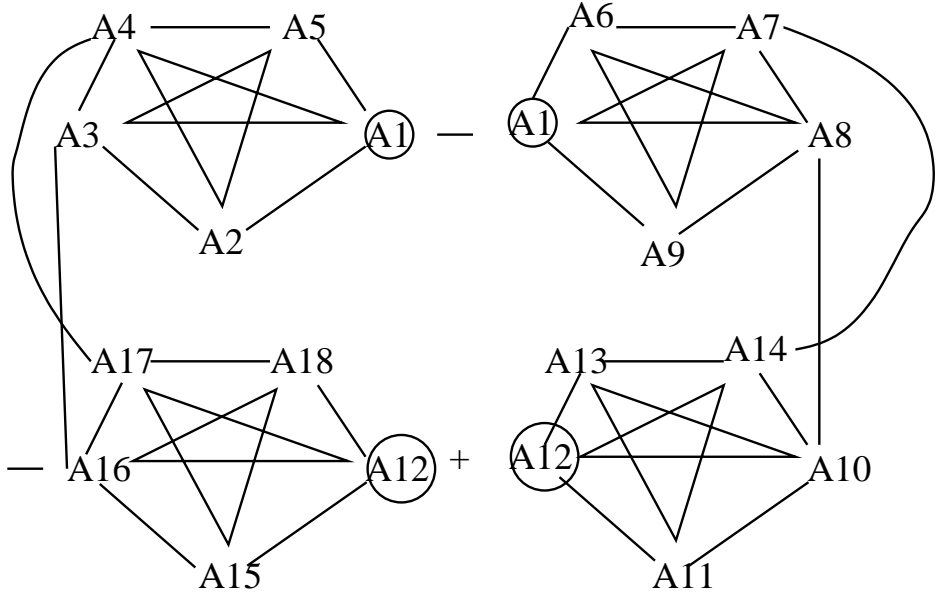}
\end{center}
\caption{Graphical representation of one term in the 5-point subleading MHV amplitude.}
\label{fig:fig1}
\end{figure}

In order to get a better understanding of the geometry in (\ref{termul}), we first express each of the $V$'s in terms of volumes of tetrahedra as
\bea
V_1&=&I(1234(5))+I(2345(1))+I(3451(2))+I(4512(3))+I(5123(4))\cr
&\equiv& A_1+A_2+A_3+A_4+A_5\cr
V_2&=&I(1234(6))+I(2346(1))+I(3461(2))+I(4612(3))+I(6123(4))\cr
&\equiv& A_1+A_6+A_7+A_8+A_9\cr
V_3&=&I(6178(4))+I(1784(6))+I(7846(1))+I(8461(7))+I(4617(8))\cr
&\equiv& A_{10}+A_{11}+A_{12}+A_{13}+A_{14}\cr
V_4&=&I(1784(5))+I(7845(1))+I(8451(7))+I(4517(8))+I(5178(4))\cr
&\equiv& A_{12}+A_{15}+A_{16}+A_{17}+A_{18}
\eea
where the tetrahedra $A_1-A_{18}$ are defined in the order they appear in the sum,
and we then represent the sum in (\ref{termul}), $V_1-V_2-V_4+V_3$ diagramatically as in Fig \ref{fig:fig1}. 
Here "vertices" are tetrahedra, and "links" are 
common faces. We did not represent all possible links, otherwise the figure would be too messy, just the ones inside each $V$, and the ones 
connecting the bottom two $V$'s with the upper two $V$'s (the signs work out such that the faces are glued together). We see that $V_1-V_2$ and 
$V_3-V_4$ have a tetrahedron cancelling out ($A_1$ and $A_{12}$), therefore they form a continous polytope. 
Since each $V$ was already a closed polytope, and all 4 $V$'s are now connected, we have a closed polytope.

All in all, we have 17 vertices appearing in $AdS_5$ space (5 original and 12 new):
\bea
&&x_1,x_2,x_3,x_4,x_5, (x_1+x_4-x_5),(x_1+x_3-x_5),(x_2+x_4-x_3),(x_1+x_4-x_3),\cr
&&(x_2+x_5-x_1),(x_3+x_5-x_1),(x_3+x_5-x_4),(x_1+x_3-x_4),\cr
&&(x_2+x_5-x_4),(x_1+x_3-x_2),(x_1+x_4-x_2),(x_3+x_5-x_2)
\eea
Out of these 17 points we have 24 tetrahedra summed over, with the 6 KK basis tree amplitudes multiplying groupings of 4, with alternating 
signs.

We next move to the $6$-point case. Now we have two objects, $A_{6;3}$ and $A_{6;4}$, but we will explicitly analyze $A_{6;3}$ only. However
we will 
see in the next subsection that the generalization to $A_{n;3}$ and then $A_{n;j}$ is straightforward. The relation to $A_{6;1}$ in 
(\ref{nonplanar}) explicitly gives
\bea
A_{6;3}(561234)&=&A_{6;1}(123456)+A_{6;1}(123546)+A_{6;1}(125346)+A_{6;1}(152346)\cr
&&+A_{6;1}(512346)+A_{6;1}(234156)+A_{6;1}(234516)+A_{6;1}(235416)\cr
&&+A_{6;1}(253416)+A_{6;1}(523416)+A_{6;1}(341256)+A_{6;1}(341526)\cr
&&+A_{6;1}(345126)+A_{6;1}(354126)+A_{6;1}(534126)+A_{6;1}(412356)\cr
&&+A_{6;1}(412536)+A_{6;1}(415236)+A_{6;1}(451236)+A_{6;1}(541236)\cr
&&\label{a6361}
\eea
Using the KK relations (\ref{KK}) and also the inversion formula
\be
A_n(12..n)=(-1)^nA_n(n...21)
\ee
we then obtain
\bea
A_{6;3}(561234)&=&
A_6(123456)[(M_6(123456)-M_6(512346))+(M_6(541236)-M_6(412356))]\cr
&&+A_6(123546)[(M_6(123546)-M_6(412356))+(M_6(451236)-M_6(512346))]\cr
&&+A_6(125346)[(M_6(125346)-M_6(412536))+(M_6(451236)-M_6(512346))]\cr
&&+A_6(152346)[(M_6(152346)-M_6(415236))+(M_6(451236)-M_6(512346))]\cr
&&+A_6(154326)[(M_6(234516)-M_6(345126))+(M_6(341526)-M_6(234156))]\cr
&&+A_6(145326)[(M_6(235416)-M_6(354126))+(M_6(341526)-M_6(234156))]\cr
&&+A_6(143526)[(M_6(253416)-M_6(534126))+(M_6(341526)-M_6(234156))]\cr
&&+A_6(143526)[(M_6(523416)-M_6(234156))+(M_6(341256)-M_6(534126))]\cr
&&+A_6(125436)[(M_6(341256)-M_6(412536))+(M_6(451236)-M_6(345126))]\cr
&&+A_6(124536)[(M_6(341256)-M_6(412536))+(M_6(541236)-M_6(354126))]\cr
&&+A_6(142536)[(M_6(341256)-M_6(412536))+(M_6(541236)-M_6(354126))]\cr
&&+A_6(124356)[(M_6(341256)-M_6(534126))+(M_6(541236)-M_6(412356))]\cr
&&+A_6(142356)[(M_6(341256)-M_6(534126))+(M_6(541236)-M_6(412356))]\cr
&&+A_6(152436)[(M_6(341526)-M_6(415236))+(M_6(451236)-M_6(345126))]\cr
&&+A_6(154236)[(M_6(341526)-M_6(415236))+(M_6(451236)-M_6(345126))]\cr
&&+A_6(145236)[(M_6(341526)-M_6(415236))+(M_6(541236)-M_6(354126))]\cr
&& \label{a63}
\eea
We have grouped the $M$ factors in brackets of two, containing opposite signs, for purposes to be explained below.

We note some patterns that will repeat at higher $n$. First, unlike the $n=5$ case, there are only 16 out of the $4!=24$ members of the 
KK basis present. These KK basis elements start with 1 and end with $n=6$, like all the KK basis members. Then $n-1=5$ can take any position 
along the remaining 4 sites, i.e. we have 4 groups of 4, each group being characterized by just changing the position of $n-1=5$. The 4 groups 
correspond to the permutations $234, 432, 243, 423$. We note that the permutations missing are $324$ and $342$. The pattern is that the 
allowed permutations of $234$ are as follows: the arrangement of the permuted objects $2,3,4$ inside the permutation goes towards the right from
2 to $j_{max}$ and then towards the left until $n-2=4$. Then indeed the only permutations that do not obey this rule are the excluded $324,342$.

We also note the rule that in each of the 2 () brackets for each KK basis term we have the same permutation of $n-2=4$ objects, and the remaining 
2 are flipped in between the two terms of the () bracket. This signifies that in the MHV case
the bracket makes the difference of the volumes of the boundaries 
of two $n=6$-polytopes with $n-1=5$ common vertices and the last one (corresponding to the $x_i$ in between the two momenta being flipped)
differing, which, according to the definition in \cite{Mason:2010pg} reviewed in the previous  subsection, 
means that the two polytopes are glued into a single one. For instance the first () bracket in the last line of (\ref{a63}) is 
(after using cyclicity)
\bea
&&M_6^{MHV}(415263)-M_6^{MHV}(415236)\cr
&=&V(x_1+x_4-x_5,x_1,x_2,x_2+x_6-x_5,x_3+x_6-x_5,x_3+x_1-x_5)\cr
&&-V(x_1+x_4-x_5,x_1,x_2,x_2+x_6-x_5,x_3+x_6-x_5,x_4+x_6-x_5)\cr
&&
\eea
This will also generalize. The 4 terms in the [] bracket multiplying the KK basis element
can be described, considering that $M$ is $M_n(\{\b\},1,\{\a\},n)$, by saying that $n-1=5$ can be either in $\{\a\}$ or in $\{\b\}$, and similarly 
$j_{max}$ can be in either $\{\a\}$ or $\{\b\}$, generalizing as well. The two flipped objects in (\ref{a63}) are $26,36,46,56$, i.e. 
generalizing to $2n,3n,...(n-2)n$ and 
also $(n-1)n$. Finally, the sign of $M$ is simply given by $n_\b$, 
i.e. the number of objects in the permutation $\{\b\}$, and this will also generalize.

In this subsection we have seen that the subleading amplitude is expressed as a sum over elements of the KK basis of 
tree amplitudes, whose coefficients are sums of volumes 
of closed polytopes, the sum representing also the volume of a closed (without a boundary) polytope. The $M_5$'s and $M_6$'s
were closed 3-dimensional polytopes with 5 and 6 vertices, and the coefficients of each tree amplitude was a sum of 4 terms, which split in two 
groups of 2. Each group is the difference of two polytopes with only one vertex different, forming together closed polytopes with 6 and 7 vertices 
respectively, and the two groups added up also form a closed polytope with a larger number of vertices. In the following we will generalize these 
results.

\subsection{Generalization to $n$-point MHV subleading amplitudes}

{\bf Generalization to $A_{n;3}$}. 

Before we generalize the result of the last subsection, we rewrite for $j=3$ the general formula (\ref{nonplanar}) as 
\cite{Bern:1990ux}
\be
A_{n;3}(n-1,n,1,2,...,n-2)=\sum_{\sigma\in COP^{(1...n-2)}_{n-1}\{1,...,n-1\}}A_{n-1}(\sigma(1),...,\sigma(n-1),n)\label{n3}
\ee
This is the same formula, since we can rewrite it as 
\bea
A_{n;3}(123...n)&=&\sum_{\sigma\in COP_{n-1}^{(34...n)}\{3,4...n,1\}}A_{n;1}(\sigma(3),...,\sigma(n),\sigma(1),2)\cr
&=&\sum_{\sigma\in COP_{\{ 21\},\{3...n\}}}A_{n;1}(\sigma)\label{rewrite}
\eea
The second line also has $(n-1)(n-2)$ terms ($n-1$ possible positions for 1, and once 1 is fixed, only $n-2$ positions for 2), and by periodicity of 
$A_{n;1}$ we can put 2 on the last position. Then 1 is on any other position, and the rest (3,4,...n) are cyclically ordered, i.e. we 
have exactly $COP^{(34...n)}_{n-1}\{3,4,...,n,1\}$, or the first line.

We want to understand the coefficients of the KK basis amplitudes in an expansion generalizing (\ref{a63}). They are all of the type 
$M_n(\{\b\},1,\{\a\},n)$, since we have used the KK relations (\ref{KK}) to get to the KK basis expansion. 
We then rewrite the amplitude on the right hand side of (\ref{n3}) as 
$M_n(\{\b\},1,\{\a\},n)$, where $n-1$ is either in $\{\a\}$ or in $\{\b\}$ (since $n-1$ is not part of the cyclic permutations), and 
otherwise $\{\a\}$ contains $2,3,...,k$ and $\{\b\}$ contains $k+1,...,n-2$. 

Then when using the KK relations (\ref{KK}), from 
$\{\a\}$ and $\{\b\}$ we form the permutation $\{\sigma\}_i$ which contains $\{\a\}$ and $\{\b^T\}$, keeping the ordering, i.e.
in the KK basis amplitude we have $A(1,\{\sigma\},n)$, where if we extract the $n-1$, $\{\a\}=2,...,j_{max}$ is ordered, i.e. it goes from 
left to right in the permutation, and then $\{\b\}=j_{max}+1,...,n-2$ is transposed and still ordered, i.e. it goes from right to left. 
That is exactly the general rule for the KK basis members that we extrapolated from the $n=6$ case before. But then we note that 
the same $j_{max}$ (extracted from the resulting KK basis member) is obtained from either $k$ or $k+1$. That means that there are 
exactly 4 terms corresponding to the same KK basis member, corresponding to both $j_{max}$ and $n-1$ belonging to either $\{\a\}$ or $\{\b\}$. 

The sign of the terms is obtained from the sign in the KK relations (\ref{KK}) (besides that sign, there are only plus signs multiplying  
terms), i.e. $(-1)^{n_\b}$, where here $\{\b\}$ refers to the individual $M_n(\{\b\},1,\{\a\},n)$ term. We now look in more detail at the 
4 terms that multiply the KK basis amplitude. It is easy to understand that $j_{max}$ belonging to either $\{\a\}$ or $\{\b\}$ means 
that in $M_n(\{\b\},1,\{\a\},n)$ we either have $j_{max}$ at the end of $\{\a\}$, or at the beginning of $\{\b\}$, i.e. we have 
a flip of $j_{max}$ $n$ vs. $n$ $j_{max}$ in between terms with different signs, since different $n_\b$ (with or without $j_{max}$). 
The exception is when actually $(n-1)$ is at the end of $\{\a\}$ and not $j_{max}$, which can then change into the first in $\{\b\}$ 
(in order to obtain the same KK basis member) in which case the same flip is now $(n-1)n$ vs. $n(n-1)$, and the same relative minus
sign applies. 

Since the pair in the difference  in the () bracket multiplying KK basis members
has the same $n-2$ permutation, and the remaining two terms are flipped, as in the $n=6$ case we have for MHV 
the difference of two $n-$polytopes with a common $n-1$-polytope. Thus we have finished proving all the generalizations we have mentioned 
in the previous subsection. 

For completeness, we will also find the total number of terms in $A_{5;3}$. In $M_n(\{\b\},1,\{\a\},n)$ we have the permutation 
$(COP_{n-1}^{(1...(n-2))}\{1,...,n-1\},n)$, but then we use the KK basis to sum over $\{\sigma\}_i\in OP(\{\a\},\{\b\})$, so we have to count these 
last permutations. So we choose $n_\a$ terms out of $n_\a+n_\b=n-2$, corresponding to choosing the spots where we put the $\{\a\}$ 
objects inside $\{\sigma\}_i$, which is the binomial coefficient $((n-2);n_\a)$. But were it not for $n-1$, $\{\a\}$ would be fixed to 
be just $2,...,n_\a+1$. As it is, we can have $n-1$ be one of the $n_\a$ terms, or one of the $n_\b=n-2-n_\a$ terms, giving finally
\be
\left(\sum_{n_\a=1}^{n-2}n_\a+\sum_{n_\a=0}^{n-3}(n-2-n_\a)\right)\frac{(n-2)!}{n_\a!(n-2-n_\a)!}
\ee
However, we can easily see that the second sum equals the first, by redefining $n-2-n_\a\equiv n_\a'$, so that we have
\be
2\sum_{n_\a=1}^{n-2}n_\a\frac{(n-2)!}{n_\a!(n-2-n_\a)!}\label{nrterms}
\ee
total number of terms, and $1/4$ as much KK basis members. This counting of KK basis members matches for n=5 and n=6, obtaining 24/4=6 and 
64/4=16 terms respectively. We note that in general, the number of KK basis terms appearing is much smaller than the total number, since by 
the Sterling formula, $n!\sim (n/e)^n$, so that (\ref{nrterms}) estimates to  
\be
\sim \sum_{n_\a=1}^{n-2}\frac{(n-2)^{n-2}}{(n-2-n_\a)^{n-2-n_\a}(n_\a-1)^{n_\a-1}}\label{sterling}
\ee
instead of $(n-2)!\sim ((n-2)/e)^{n-2}$, which is much larger than (\ref{sterling}).

Finally, the formula for $A_{n;3}^{MHV}$ can be written as 
\bea
A_{n;3}^{MHV}(n-1,n,1,2,....,n-2)&=&\sum_{\{\sigma\}_i\in OP(\{\a\},\{\b^T\})|_{j_{max}}}A_n^{MHV}(1,\{\sigma\}_i,n)\cr
&&\times \sum_{n-1\in \{\a\},\{\b\};j_{max}\in\{\a\},\{\b\}}(-)^{n_\b}M_n^{MHV}(\{\b\},1,\{\a\},n)\cr
&&\label{an3}
\eea
with $M_n^{MHV}$ being the volume of a closed polytope, and pairs of opposite sign $M_n$'s adding up to another closed polytope, as explained above.

{\bf Generalization to $A_{n;j}$.} 

The generalization to $A_{n;j}$ is now also straightforward, though it becomes more involved. We rewrite the formula 
for $A_{n;j}$ as a function of $A_{n;1}$ as
\be
A_{n;j}(1...n)=(-1)^{j-1}\sum_{\sigma\in COP_{\{j-1,...,1\},\{j,...,n\}}}A_{n;1}(\sigma)=\sum_{\sigma\in COP_{\{j,...,n\},\{j-2,...,1\}}}
A_{n;1}(\sigma, j-1)\label{anj}
\ee
where in the second equality we have used the cyclic property of $A_{n;1}(\sigma)$. Then as before we write $A_{n;1}(...)\equiv A_n(...)M_n(...)$, and 
we rewrite the $A_n(...)$'s in terms of the KK basis obtaining
\be
A_n(n-j+2,...,n,1,...,n-j+1)=\sum_{\sigma\in COP_{\{1,...,n-j+1\},\{n-1,...,n-j+2\}}}A_n(\sigma, n)\label{intermediate}
\ee
Then we get $M_n(\{\b\},1,\{\a\},n)$, where $\{\a\}=2,...,j_{max}$ with or without $\{n-1,...,n-j+2\}$. At the end of the 
permutation $\sigma$ in $A(\sigma, n)$, just before $n$, we can have either $j_{max}$, OR one of $\{n-1,...,n-j+2\}$, i.e. one of ($j-2$ terms).

Then the KK basis elements that we get are of a special type: If we take out $n-1,...,n-j+2$ from the amplitude, 
then the situation should be like the one for $n=3$, namely in the remaining permutation we go from 1 to a $j_{max}$ towards the right, and then towards
the left. But moreover, in $n-1,..., n-j+2$ we also have some ordering: some of them are in $\{\a\}$, some in $\{\b^T\}$, which means that 
$n-1,...,l_{max}+1$ is cyclic (i.e., towards the right), and $n-j+2,...,l_{max}$ is also cyclic (i.e., we change the direction of the cyclicity
at $l_{max}$). 

Then the number of terms multiplying a KK basis member is even, corresponding to having $j_{max}$ in $\{\a\}$ or $\{\b\}$ 
and any number of the $j-2$ terms $\{n-1,...,n-j+2\}$ in $\{\a\}$ and the rest in $\{\b\}$. They come in pairs, the pairs corresponding to 
$j_{max}$ being just before $n$ or just after, or otherwise one of the $\{n-1,...,n-j+2\}$ being either just before, or just after $n$, and 
the pairs as before having different sign.  
The sign of the terms is then simply $(-1)^{j-1+n_\b}$. In terms of polytopes, the two terms of different sign
correspond as before to polytopes with only a vertex differing between them, which means they again add up to another polytope with one more vertex.

The final formula for $A_{n;j}$ is:
\bea
&&A_{n;j}^{MHV}(n-j+2,...,n,1,...,n-j+1)\cr
&=&\sum_{\{\sigma\}_i\in OP(\{\a\},\{\b^T\})|_{j_{max}\in \{1,...,n-j+1\},l_{max}\in\{n-j+2,...,n-1\}}}
A_n^{MHV}(1,\{\sigma\}_i,n)\cr
&&\times \sum_{\{n-1,...,n-j+2\}\in \{\a\},\{\b\};j_{max}\in\{\a\},\{\b\}}(-)^{n_\b+j-1}M_n^{MHV}(\{\b\},1,\{\a\},n)\cr
&&\label{anjfinal}
\eea
where the complicated notation is explained above, and again we have pairs of $M_n^{MHV}$'s of different signs adding up to give other 
closed polytopes (of $n+1$ vertices).

\subsection{Leading IR singularities of $A_{n,j}$}

At one-loop, (\ref{5pointKK}), (\ref{a63}), (\ref{an3}) and (\ref{anjfinal}) show that the amplitudes $M_n$ come in alternating pairs. Each of 
these $M_n$ has leading IR singularity $1/\epsilon^2+{\cal O}(1/\epsilon)$, and therefore at one-loop $A_{n,j}$ has only a $1/\epsilon$ IR 
singularity. 

In \cite{Naculich:2008ys} and \cite{Naculich:2009cv} it was shown that the leading IR singularity of the most subleading color amplitude for 
gluon-gluon scatteing in ${\cal N}=4$ SYM at L-loops is $1/\epsilon^L$, which is compatible and complementary to the results of this paper. 
For amplitudes which are not the most subleading, the leading IR singularity is $1/\epsilon^{L+k}$ for $k=1,...,L$, where $1/\epsilon^{2L}$ corresponds
to the single trace, i.e. planar terms.

\section{Twistor methods for $N^kMHV$ subleading amplitudes}

\subsection{Review of twistor methods for leading amplitudes}

Writing the amplitudes and superamplitudes in twistor space has proved to be a very useful tool over the last years. 
In the previous section, we made use of the dual space $x_i$ which trivializes (identically solves) the momentum conservation condition 
$\sum_{i=1}^n p_i=0$ by $p_i=x_i-x_{i+1}$. On the other hand, for amplitudes of massless particles, we are interested in trivializing the 
lightlike condition $p^2=0$ by the use of helicity spinors $p_{\a\dot\a}=\lambda_\a\tilde \lambda_{\dot\a}$. We can use this to write 
twistor space representations for leading singularities of amplitudes. 

The leading singularities of an amplitude are the discontinuities of the amplitude over the singularities where we put a maximum number of propagators 
on-shell, as explained in \cite{ArkaniHamed:2009dn}, where a conjecture for these leading singularities was proposed. The conjecture is that the 
same integral formula can be written for leading singularities of all the  (color-ordered, planar, i.e. leading) loop amplitudes and for the full 
(color-ordered) tree amplitude, where only integration contours are different. Moreover, the formalism naturally encompases superamplitudes.

The MHV tree-level color-ordered superamplitudes are given by the Nair formula \cite{Nair:1988bq}, 
a supersymmetric generalization of the Parke-Taylor formula \cite{Parke:1986gb,Berends:1987me}, 
\be
{\cal A}_{n,2}(12...n)=\frac{\delta^4(\sum_{i=1}^n\lambda_i\tilde\lambda_i)\delta^8(\sum_{i=1}^n\lambda_i\tilde \eta^i)}{\langle 12\rangle\langle
23\rangle...\langle n-1,n\rangle \langle n,1\rangle}\label{Nair}
\ee
where as usual $\langle ij\rangle\equiv \epsilon^{\a\b}\lambda_\a^{(i)}\lambda_\b^{(j)}$, $\tilde\eta$ is a spinor with an index $I=1,...,4$ for 
supersymmetries suppressed, and the $2$ in $A_{n,2}$ refers to R-charge, since 
the $N^kMHV$ amplitude has $m=k+2$ R-charge.

Then the leading singularities of the R-charge $m$ superamplitudes are given by \cite{ArkaniHamed:2010gh}
\be
{\cal L}_{n,m}=\int\frac{d^{nm}C_{\mu i}}{Vol(Gl(m))}\prod_{\mu=1}^m\frac{\delta^2(\sum_{i=1}^nC_{\mu i}\tilde \lambda_i)\delta^{0|4}(\sum_{i=1}^n
C_{\mu i}\tilde\eta_i)}{(12...m)(23...m+1)...(n12...m-1)}\int\prod_{\mu=1}^m d^2\rho_\mu\prod_{i=1}^n\delta^2(C_{\mu i}\rho_\mu-\lambda_i)
\label{twistorint}
\ee

Here $(12...m)$, etc. are Plucker coordinates on the Grassmanian $G(m,n)$, i.e. determinants of $m\times m$ minors of the matrix 
\be
\begin{pmatrix}C_{11}&C_{12}&...& C_{1n}\\C_{21}&C_{22}&...&C_{2n}\\..&..&...&..\\C_{m1}&C_{m2}&...&C_{mn}\end{pmatrix}
\ee
After doing the integration over the delta functions, one is left with an $(m-2)[(n-4)-(m-2)]$ dimensional integral to be done. 
Note that in the MHV case ($m=2$), there
is no integral left to be done, and the formula for the leading singularity reduces to 
\be
{\cal A}_{n,2}=
\int \frac{d^{2n}C_{\mu i}}{Vol(Gl(2))}\frac{\prod_{\mu=1}^2\delta^2(\sum_{i=1}^nC_{\mu i}\tilde \lambda_i)\delta^{0|4}(\sum_{i=1}^nC_{\mu
 i}\tilde\eta_i)}{(12)(23)...(n1)}\int \prod_{\mu=1}^2 d^2\rho_\mu\prod_{i=1}^n\delta^2(C_{\mu i}\rho_i-\lambda_i)\label{mhvtwistor}
\ee
which gives nothing but the tree MHV superamplitude (\ref{Nair}). 

Taking a Fourier transform over $\lambda$, 
\be
\tilde{\cal L}_{n,m}\equiv \int\prod_{i=1}^nd^2\lambda_ie^{i<\lambda_i,\mu_i>}{\cal L}_{n,m}
\ee
and defining
\be
{\cal W}_i=\begin{pmatrix}\tilde \lambda_i\\\mu_i\\\tilde\eta_i\end{pmatrix}\in C^{4|4}
\ee
where the physical configuration actually lives in dual supertwistor space $CP^{3|4}\in C^{4|4}$. The leading singularity of the (color-ordered, 
planar, i.e. leading) amplitude is written in dual supertwistor space as 
\be
\tilde {\cal L}_{n,m}=\int \frac{d^{nm}C_{\mu i}}{Vol(Gl(2))}\frac{\prod_{\mu=1}^m\delta^{4|4}(\sum_{i=1}^n C_{\mu i}{\cal W}_i)}{(12...m)(23...m+1)...
(n12...m)}
\ee
Taking the Fourier transform over the whole ${\cal W}_i$, we obtain the leading singularity of the amplitude in  twistor space parametrized 
by $Z_i$,
\bea
{\tilde{\tilde {\cal L}}}_{n,m}&\equiv &\int \prod_{i=1}^n d{\cal W}_ie^{Z_i\cdot{\cal W}_i}\tilde {\cal L}_{n,m}({\cal W})\cr
&=&\int \prod d\xi_\mu^{4|4}\int \frac{d^{nm}C_{\mu i}}{(12...m)...(n12...m)}\prod_{i=1}^n\delta^{4|4}(Z_i-\sum_\mu C_{\mu i}\xi_\mu)
\eea

An even more powerful way to write the superamplitude is using Hodges' momentum twistors \cite{Hodges:2009hk,Hodges:2010kq}. 
Indeed, it was shown in \cite{ArkaniHamed:2009vw} that the form conjectured in \cite{ArkaniHamed:2009dn} can be written in terms of momentum 
twistors. These are defined as follows. For each of the dual coordinates $x_i$ we associate 
an 
\be
\begin{pmatrix} \lambda\\ \mu\end{pmatrix}=Z\in C^4
\ee
defined by 
\be
(\mu_i)_{\dot\a}=(x_i)_{\a\dot\a}\lambda_i^\a
\ee
but since the data we are interested in is $x_{\a\dot\a}$, which is invariant under rescalings of $Z$, $Z$ lives in $CP^3$= twistor space, more 
specifically, defines a $CP^1\in CP^3$. Thus a null ray in (dual) spacetime corresponds to a point in twistor space, but the (dual) spacetime point 
$x_{\a\dot\a}$ defines a $CP^1$ line in twistor space. Given two points in spacetime $X_{\a\dot\a}$ and $Y_{\a\dot\a}$, they are null 
separated $\Leftrightarrow$ the 2 $CP^1$'s corresponding to them intersect on a single point. Therefore in {\bf momentum - twistor space}, 
corresponding to our $x_i$'s with $p_i=x_i-x_{i+1}$, we have polygons made up of $CP^1$ lines for $x_i$, intersecting over 
vertices = momentum twistors $Z_i$, thus ensuring both the null separation and momentum conservation of the original momenta.
Reversely, given an {\em arbitrary} set of $n$ twistors $Z_i$, we can construct $CP^1$'s connecting them, thus deriving $n$ null momenta satisfying 
momentum conservation.

In terms of super-momentum twistors $Z_i$, the leading singularity of the (color-ordered, planar, i.e. leading) $N^kMHV$ amplitude is 
\cite{ArkaniHamed:2009vw,ArkaniHamed:2010gh}
\be
{\cal L}_{n,m}=\frac{\delta^4(\sum\lambda\tilde\lambda)\delta^8(\sum\lambda\tilde\eta)}{\langle12\rangle\langle 23\rangle...\langle
n1\rangle}\int\frac{d^{nk}{\cal D}}{Vol(Gl(2))}\frac{\prod_{\mu=1}^{k}\delta^{4|4}(\sum_{i=1}^n{\cal D}_{\mu 
i}Z_i)}{(12...k)(23...k+1)...(n12k-1)}={\cal L}_{n,2}\times {\cal R}_{n,k}\label{yangian}
\ee
where $k=m-2$. The prefactor ${\cal L }_{n,2}$ is the tree MHV amplitude (\ref{Nair}), 
and the integral ${\cal R}_{n,k}={\cal R}_{n,m-2}$ is Yangian invariant. 
This object is dual conformal {\em covariant}, only ${\cal R}_{n,k}$ being DCI, and the tree amplitude is covariant.

As an aside, we note that one can rewrite the full loop amplitudes corresponding to Feynman diagrams in momentum twistor space, 
as {\em integrals over an integrand}, and at the level of this integrand, we can decompose it in the integrands of boxes and pentagons
\cite{ArkaniHamed:2010gh}.
The same momentum-twistor space integrand for the amplitude can be decomposed into integrands of scalar integrals with {\em unit chiral 
leading singularities}, which are found to be octagon integrals in general \cite{ArkaniHamed:2010gh}.

\subsection{Application to subleading $N^kMHV$ amplitudes}

We now return to the issue of the leading singularities and their connection with the amplitudes. They are obtained as residues of 
singularities found by putting a maximal number of loop lines on-shell. We will be interested in the 1-loop case, for which that maximal 
number is 4.  Thus for the leading singularity, cutting 4 loop propagators, we are left with a product of 4 tree amplitudes, which in the MHV 
case ($m=2$) is just a representation of the MHV tree amplitude via recurrence relations, $R\rightarrow\sum M^{(0)}_a M^{(0)}_b M^{(0)}_c M^{(d)}_d$.
This gives a simple test of the formula (\ref{yangian}), which for MHV ($m=2$) reduces to just the MHV tree amplitude, coinciding with the previous 
result at one-loop. 

The one-loop amplitudes of ${\cal N}=4$ SYM can be reduced to just boxes via the van Neerven and Vermaseren procedure
(for a general gauge theory, we need also triangles and bubbles), with some coefficients. 
At one-loop, the leading singularities also coincide with the coefficients of these box functions \cite{ArkaniHamed:2009dn}, though at 
higher loops, this is 
more complicated. For one-loop MHV, the coefficients of the boxes are known to be just the MHV tree amplitudes, agreeing with the result above. 

At one-loop, cutting the 4 loop propagators from the box functions in order to obtain the leading singularities gives rise to the 
Schubert problem (1850) for finding the residues: put
\be
(x-x_i)^2=0;\;\;\; (x-x_k)^2=0;\;\;\; (x-x_l)^2=0;\;\;\; (x-x_s)^2=0
\ee
corresponding to the sought-for residues, or leading singularities. This can be mapped to the problem of having 4 lines in twistor space $CP^3$, 
and finding all the lines that intersect all of them. There are exactly two 
solutions to this problem for generic 4 lines in $CP^3$, therefore there are exactly two residues, or leading singularities. 

Finally, putting everything together, the planar (leading) color-ordered $N^kMHV$ amplitude is a sum of permutations of boxes with coefficients equal 
to the leading singularities,
\be
A_{n;1}(1...n)=\sum_\sigma {\cal L}_{n,k}(\sigma) I_{n;4}(\sigma)=\sum A^{MHV}_n(\sigma)R_{n;k}(\sigma)I_{n;4}(\sigma)
\ee
where $I_{n;4}$ are boxes. At 6-points, 
the permutations $\sigma$ combine such that we can organize the sum as a sum over cyclic permutations, with several boxes having the same coefficient
\cite{ArkaniHamed:2009dn}. 
For this coefficient we can factorize the tree MHV amplitude, which is cyclically invariant, so that it appears as a common factor 
\be
A_{6;1}(1...6)=A_6^{MHV}(1...6)\sum_{\lambda= cyclic} R_{6;k}(\lambda)\sum_{\sigma/\lambda}I_{6;4}(\sigma)\label{a6twistor}
\ee
At higher $n$-point, the situation is slightly more complicated. The box diagrams are ordered in groups that can be cyclically permuted, 
for each group having a given formula for the residue, but unlike 6-point, the residue is not universal for all the 
groups \cite{ArkaniHamed:2009dn}. However, all 
the diagrams have still the external legs in the original order, which means, since the MHV tree amplitude
is cyclically invariant, that we can again factorize the MHV tree amplitude, obtaining for planar $N^kMHV$ amplitudes
\bea
A_{n;1}(1...n)&=&A_n^{MHV}(1...n)\sum_{groups\; of\; diagrams}\sum_{\lambda= cyclic} R_{n;k}(\lambda)\sum_{\sigma/\lambda}I_{n;4}(\sigma)\cr
&\equiv&A_n^{MHV}(1...n)M_{n;k}(1...n)\label{mnk}
\eea
which implicitly defines $M_{n,k}$. 

We now finally note that we have the same formula for $A_{n;1}(1...n)$ in terms of $A_n^{MHV}$ and $M_{n;k}$ from the previous section 
on polytopes, so we can apply the same calculations we used to obtain the $MHV$ $A_{n;j}$ in terms of $A_{n;1}$ in section 2, but now just change the 
definition of $M_{n;k}$ as in (\ref{mnk}), thus also drop the polytope interpretation of $M_{n,k}$. 
But otherwise the same (\ref{anjfinal}) found in the MHV case holds in the general $N^kMHV$ case as well, as can be seen from  
(\ref{rewrite}), (\ref{an3}) and (\ref{intermediate}). A trivial case is $n=5$, $NMHV$, which is given by a single R-invariant
\cite{Drummond:2008bq}.

\section{The 6-point subleading NMHV amplitudes}

The 6-point subleading MHV amplitude was calculated using polytope methods in section 2.2. We now consider the NMHV 
amplitudes in detail. In the previous section we found formulas for the superamplitude using twistor methods. However, we have seen that we can 
write boxes in terms of tetrahedra and polytopes. The 6-point NMHV can be written explicitly in terms of boxes, so we anticipate a 
polytope interpretation here too.

\subsection{Polytope picture for leading 6-point NMHV amplitudes}

The leading (planar) gluon amplitudes $A_{6;1}^{NMHV}$ were found in \cite{Bern:1994cg}. For the split-helicity configuration 
\be
A_{6;1}^{NMHV}(1^+2^+3^+4^-5^-6^-)=\frac{c_\Gamma}{2}(B_1 W_6^{(1)}+B_2 W_6^{(2)}+B_3W_6^{(3)})\label{1lnmhv}
\ee
where $W_6^{(i)}$ are cyclic permutations of $W_6^{(1)}$, and $W_6^{(i+3)}\equiv W_6^{(i)}$. 
The $W_6^{(i)}$ are given in terms of box functions by 
\be
W_6^{(i)}=F_{6:i}^{1m}+F_{6:i+3}^{1m}+F_{6:2;i+1}^{2mh}+F_{6:2;i+4}^{2mh}\label{ws}
\ee
and the $F$'s are are dimensionless boxes, related to the usual box integrals $I$ by
\bea
&&F_{n;i}^{1m}=-\frac{t_{i-3}^{[2]}t_{i-2}^{[2]}}{r_\Gamma}I^{1m}_{4:i}\cr
&& F_{n:r;i}^{2mh}=-\frac{t_{i-2}^{[2]}t_{i-1}^{[r+1]}}{2r_\Gamma}I_{4:r;i}^{2mh}\label{fis}
\eea
where $t_i^{[r]}=(p_i+...+p_{i+r-1})^2$. The box integral $I_{4:i}^{1m}$ has massless particles $i-3,i-2,i-1$ at 3 consecutive vertices and 
$i,...$ at the other one,
while $I_{4:r;i}^{2mh}$ has massless particles $i-2,i-1$ at two consecutive vertices, $i,...,i+r-1$ at the next and $i+r,...$ at the last. 
In the above
\bea
r_\Gamma&=&\frac{\Gamma(1+\epsilon)\Gamma^2(1-\epsilon)}{\Gamma(1-2\epsilon)}\simeq 1+{\cal O}(\epsilon)\cr
c_\Gamma&=&\frac{r_\Gamma}{(4\pi)^{2-\epsilon}}
\eea
The other helicity configurations are 
\bea
&&A_{6;1}^{NMHV}(1^+2^+3^-4^+5^-6^-)=\frac{c_\Gamma}{2}(D_1 W_6^{(1)}+D_2 W_6^{(2)}+D_3W_6^{(3)})\cr
&&A_{6;1}^{NMHV}(1^+2^-3^+4^-5^+6^-)=\frac{c_\Gamma}{2}(G_1 W_6^{(1)}+G_2 W_6^{(2)}+G_3W_6^{(3)})
\eea
where the spin factors $B_i,D_i$ and $G_i$ are given in \cite{Bern:1994cg,Kosower:2010yk}. 

The tree amplitudes are 
\bea
A(1^+2^+3^+4^-5^-6^-)&=&\frac{1}{2}[B_1+B_2+B_3]\cr
A(1^+2^+3^-4^+5^-6^-)&=&\frac{1}{2}[D_1+D_2+D_3]\cr
A(1^+2^-3^+4^-5^+6^-)&=&\frac{1}{2}[G_1+G_2+G_3] 
\eea

We have seen that the dimensionless boxes can be written as polytopes, for instance $F_{6:1}^{1m}=F_{1m}(123456)$ is the polytope previously denoted by
$I(x_4,x_5,x_6,x_1(x_2,x_3))$, that we will now simply denote by $(4561(23))$, meaning $x_2$ and $x_3$ are omitted in writing the vertices. Then explicitly writing the $W$'s as polytopes in (\ref{ws}), we get 
\bea
W_6^{(1)}&=&(4561(23))+(1234(56))+(12(3)4(5)6)+(45(6)1(2)3)\cr
&\equiv&A_1+A_3+A_2+A_4\cr
W_6^{(2)}&=&(5612(34))+(2345(61))+(23(4)5(6)1)+(56(1)2(3)4)\cr
&\equiv&A_5+A_7+A_6+A_8\cr
W_6^{(3)}&=&(6123(45))+(3456(12))+(34(5)6(1)2)+(61(2)3(4)5)\cr
&\equiv&A_9+A_{11}+A_{10}+A_{12}
\eea
where the $A$'s are tetrahedra defined in the order they appear in the $W_6^{(i)}$ above.

For comparison, we also write the 6-point MHV amplitude,
\bea
A_{6;1}^{MHV}(123456)&=&A(123456)[(12(3)45(6))+(23(4)56(1))+(34(5)61(2))\cr
&&+(1234(56))+(2345(61))+(3456(12))\cr
&&+(4561(23))+(5612(34))+(6123(45))]\cr
&=&A(123456)[A_{13}+A_{14}+A_{15}\cr
&&+A_3+A_7+A_{11}+A_1+A_5+A_9]
\eea
where again the various $A$'s are defined in the order they appear. 
Thus both $A_{6;1}^{MHV}$ and $W_6^{(i)}$ can be represented in terms of polytopes, and $A_{6;1}^{NMHV}$ is given by $W_6^{(i)}$ multiplied 
by spin coefficients. The polytopes are obtained by gluing tetrahedra along common faces of opposite orientation. A graphical representation 
of the resulting polytopes can be obtained by drawing the tetrahedra $A_1$ to $A_{15}$ as "vertices", and the lines between them representing 
a common face. Then $A_{6;1}^{MHV}$ is represented schematically as 
\bea
A_{6;1}^{MHV}(123456)/A(123456)&=&\nonumber
\eea

\begin{figure}[h]
\begin{center}
\includegraphics{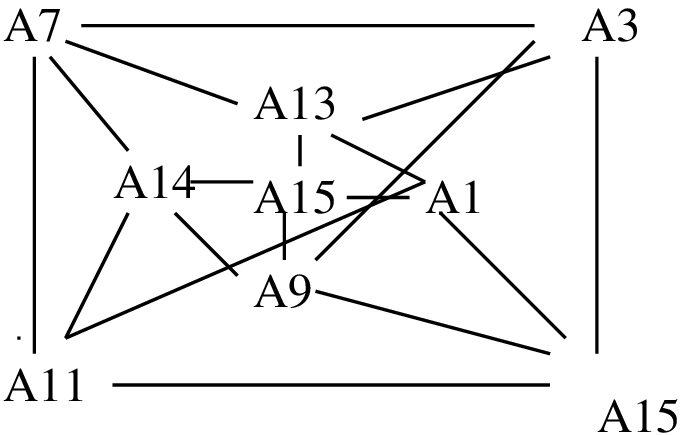}
\end{center}
\caption{.}
\label{fig:fig2}
\end{figure}

and $W_6^{(i)}$ can be represented as 
\bea
W_6^{(1)}&=&A_1----A_2\cr
&& |\hspace{2cm} |\cr
&& |\hspace{2cm} |\cr
&&A_4----A_3\cr
&&\cr
W_6^{(2)}&=&A_5----A_6\cr
&& |\hspace{2cm} |\cr
&& |\hspace{2cm} |\cr
&&A_8----A_7\cr
&&\cr
W_6^{(3)}&=&A_9----A_{10}\cr
&& |\hspace{2cm} |\cr
&& |\hspace{2cm} |\cr
&&A_{12}----A_{11}\cr
&&
\eea

We then see that the $W_6^{(i)}$'s have a polytope interpretation, though of course the coefficientts $B,D,G$ do not.

\subsection{Subleading NMHV 6-point amplitude}

We can attempt to use the polytope representation from the previous subsection for the subleading amplitudes. 
Equation (\ref{a6361}) is still valid even for NMHV, but it is not clear if we can go further than that using (\ref{1lnmhv}), 
since now not only $W_i$, but also $B_i$ 
depend on  the order of the momenta (123456), so they would need to be permuted accordingly to obtain $A_{6;3}$. 
Also note that when we permute we change $B_i$'s to $D_i$'s and $G_i$'s, 
since we also change the order of helicities in the amplitude. This is to be contrasted with the previous cases (MHV, and non-MHV in terms of 
twistors), when the helicity information was exclusively contained in the overall $A_{tree}^{MHV}(123456)$ factor, allowing the use of 
simple combinatorics. 

We also note that, because of the general properties of the color decomposition ($A_{6;1}$ multiplies a cyclic trace), $A_{6;1}^{NMHV}$ must 
be cyclically invariant, though that is not manifest from the explicit formulas we wrote. Given the cyclical invariance, we can again 
write (\ref{anj}), which for $n=6$ gives
\bea
&&A_{6;3}(1...6)=\sum_{\sigma\in COP_5^{(3456)}}A_{6;1}(\sigma,2)\cr
&&A_{6;4}(1...6)=\sum_{\sigma\in COP_{\{456\},\{2,1\}}}A_{6;1}(\sigma,3)
\eea
which gives after a shift, and considering that the $\{2,1\}$ is superflous in the second relation (up to cyclic means that the order of 1,2 is 
irrelevant, and then $COP\{2,1\}$ is irrelevant),
\bea
&&A_{6;3}(561234)=\sum_{\sigma\in COP_5^{(1234)}}A_{6;1}(\sigma,6)\cr
&&A_{6;4}(456123)=\sum_{\sigma\in COP_5^{(123)}}A_{6;1}(\sigma,6)
\eea
But this is the most we can say if we use the representation (\ref{1lnmhv}) for $A_{6;1}$. 

A more useful representation is the superamplitude obtained from an explicit form of the twistor formula (\ref{a6twistor}), 
doing the twistor space integrals over 
the 1-loop NMHV contours. The result is \cite{Kosower:2010yk,Drummond:2008bq}
\bea
A_{6;1}^{(1)NMHV}(123456)&=&\frac{a}{2}A_6^{(0)MHV}(123456)[(R_{413}+R_{146})W_6^{(1)}\cr
&&+(R_{524}+R_{251})W_6^{(2)}+(R_{635}+R_{362})W_6^{(3)}]\cr
&\equiv &A_6^{(0)MHV}(123456) M_6^{(1)NMHV}(123456)\label{6nmhv}
\eea
From the $R_{n;k}$ terms in (\ref{a6twistor}), one gets the sum of DCI basic R-invariants $R_{j,j+3,j+5}$ above. 
The R-invariants themselves have a geometric interpretation in a dual momentum twistor space 
\cite{Mason:2009qx,Hodges:2009hk,Hodges:2010kq,ArkaniHamed:2010gh,ArkaniHamed:2010gg}.
Here the $R_{j,j+3,j+5}$ are given by
\bea
R_{rst}&=&-\frac{\langle s-1\; s\rangle\langle t-1\; t\rangle \delta^{(4)}(\Xi_{rst})}{x^2_{st}\langle r|x_{rt}x_{ts}|s-1\rangle\langle r|x_{rt}x_{ts}
|s\rangle\langle r|x_{rs}x_{st}|t-1\rangle \langle r|x_{rs}x_{st}|t\rangle }\cr
\Xi_{rst}&=&\sum_t^{r-1}\eta_i\langle i|x_{ts}x_{sr}|r\rangle +\sum_r^{s-1}\eta_i\langle i|x_{st}x_{tr}|r\rangle\cr
x_{st}&=&x_s-x_t=\sum_{i=s}^{t-1}p_i
\eea
and from \cite{Mason:2009qx,Drummond:2008bq}
\be
R_{r,r+2,s}=R_{r+2,s,r+1}
\ee
we have the explicit relations among the $R$ coefficients
\bea
&&R_{413}=R_{241}=R_{624}\cr
&&R_{135}=R_{352}=R_{524}\cr
&&R_{246}=R_{463}=R_{635}\cr
&&R_{146}=R_{514}=R_{351}\cr
&&R_{513}=R_{136}=R_{362}\cr
&&R_{251}=R_{625}=R_{462}
\eea
These are further constrained by the relation \cite{Mason:2009qx,Drummond:2008bq}
\be
R_{624}+R_{625}+R_{635}=R_{146}+R_{136}+R_{135}
\ee

As explained in the section 3, 
we can then perform the same combinatorics that led us to (\ref{a63}) on the superamplitude $A_{6;1}^{(1)NMHV}(123456)$, just that now we use 
the $M_6^{(1)NMHV}(123456)$ in (\ref{6nmhv}) instead of the $M_6(123456)$. 
This way we obtain a twistor representation for the subleading 6-point NMHV superamplitude. 
We should also note that the R-invariants $R_{j,j+3,j+5}$ appearing 
in (\ref{6nmhv}) have a polytope interpretation, this time in dual momentum twistor space \cite{Mason:2009qx}, while the $W_6^{(i)}$'s we saw had a 
polytope representation in $AdS_5$.

One can do similar manipulations at higher points, as already explained.
For $n=7$ $NMHV$ amplitudes, \cite{Drummond:2008bq} give results analogous to (\ref{6nmhv}), and a similar polytope interpretation follows.

\section{Conclusions}

In this paper we have extended results for the 1-loop leading color-ordered amplitudes of ${\cal N}=4$ SYM to their subleading-color counterparts. 
Specifically, the known polytope interpretation for $MHV$ amplitudes and the momentum-twistor formulas for $N^kMHV$ amplitudes were generalized.
The subleading-color amplitudes can be obtained from the leading-color ones by (\ref{nonplanar}), but one may have wondered whether the necessary 
combinatorics become prohibitive at larger order $n$, or whether the nice interpretation at leading order remains valid. 
However, we show that this extension is quite manageable. This is a relevant issue 
for possible extensions to QCD, when the subleading pieces also need to be calculated effectively. 

We have shown that by using the KK basis of $MHV$ tree amplitudes we can write a simpler general formula (\ref{anjfinal}), which one 
can use to find other properties. As a simple application we have found that the leading $1/\epsilon^2$ IR divergence of $A_{n;j}$ cancels, leaving
only a ${\cal O}(1/\epsilon)$ IR divergence for all one-loop subleading amplitudes (the same degree of IR divergence as one-loop gravitational 
scattering amplitudes). 
In the $MHV$ case we have found that the coefficients of the KK basis amplitudes can be written as sums of volumes of simple polytopes. 
In the case $A_{n;3}$ the coefficients of the KK basis reduce to the sum of two closed polytopes with two common points. In the $N^kMHV$ case
the same formula in terms of an expansion in the KK basis of tree $MHV$ amplitudes holds, with coefficients which are momentum-twistor integrals times 
boxes. 

We have made concrete application of these concepts to the case of 5-points and 6-points. 
In the $MHV$ case we have written explicitly the formulas in terms 
of the KK basis with polytope coefficients. In the $NMHV$ 6-point case we have found a polytope interpretation for the leading amplitudes, and 
for the subleading case we have written an explicit formula in terms of R-invariants obtained by doing momentum-twistor integrations. 

{\bf Acknowledgements} 

The research of H.J. Schnitzer is supported in part by the DOE under grant DE-FG02-92ER40706 and the research of H. Nastase is supported in part by 
CNPQ grant 301219/2010-9. We thank the referee of this paper for his valuable suggestions in improving and clarifying the presentation. 

\newpage

\begin{appendix}

\section{Explicit formulas for 5-point and 6-point MHV amplitudes}

The 12 simplices appearing in $A_{5;3}$ are explicitly
\bea
M_5^{MHV}(12345)&=&V(x_1,x_2,x_3,x_4,x_5)\cr
M_5^{MHV}(23145)&=&V(x_1,(x_1-x_2+x_3),(x_1-x_2+x_4),x_4,x_5)\cr
M_5^{MHV}(31245)&=&V(x_1,(x_1-x_3+x_4),(x_2-x_3+x_4),x_4,x_5)\cr
M_5^{MHV}(12435)&=&V(x_1,x_2,x_,(x_3-x_4+x_5), x_5)\cr
M_5^{MHV}(14235)&=&V(x_1,x_2,(x_2-x_1+x_5),(x_3-x_4+x_5),x_5)\cr
M_5^{MHV}(31425)&=&V((x_1-x_2+x_3),(x_1-x_2+x_4),x_4,x_5,(x_5-x_2+x_3))\cr
M_5^{MHV}(34125)&=&V(x_3,(x_1+x_3-x_5),(x_1+x_4-x_5),x_1,x_2)\cr
M_5^{MHV}(41235)&=&V((x_4-x_5+x_1),x_1,x_2,x_3,x_4)\cr
M_5^{MHV}(43125)&=&V(x_5,x_1,(x_1-x_4+x_5),(x_1-x_3+x_5),(x_2-x_3+x_5))\cr
M_5^{MHV}(42315)&=&V((x_4-x_5+x_1),x_1,(x_1-x_2+x_3),(x_1-x_2+x_4),x_4)\cr
M_5^{MHV}(23415)&=&V(x_2,x_3,x_4,x_5,(x_5-x_1+x_2))\cr
M_5^{MHV}(24315)&=&V(x_2,x_3,(x_3-x_4+x_5),x_5,(x_5-x_1+x_2))
\eea

The coefficients of the expansion in KK basis elements are 
\bea
A_5^{MHV}(12345)&:&M_5^{MHV}(12345)-M_5^{MHV}(41235)+M_5^{MHV}(43125)-M_5^{MHV}(31245)\cr
&=&V(x_1,x_2,x_3,x_4,x_5)-V((x_4-x_5+x_1),x_1,x_2,x_3,x_4)\cr
&&+V(x_4,(x_1+x_4-x_5),x_1,(x_1-x_3+x_4),(x_2-x_3+x_4))\cr
&&-V(x_1,(x_1-x_3+x_4),(x_2-x_3+x_4),x_4,x_5)\cr
A_5^{MHV}(12435)&:&M_5^{MHV}(12435)-M_5^{MHV}(31245)+M_5^{MHV}(34125)-M_5^{MHV}(41235)\cr
&=&V(x_1,x_2,x_3,(x_3+x_5-x_4),x_5)\cr
&&-V((x_1-x_4+x_3),x_1,x_2,x_3,(x_5-x_4+x_3))\cr
&&+V(x_3,(x_1+x_3-x_5),(x_1+x_4-x_5),x_1,x_2)\cr
&&-V((x_1+x_4-x_5),x_1,x_2,x_3,x_4)\cr
A_5^{MHV}(14235)&:&M_5^{MHV}(14235)-M_5^{MHV}(31425)+M_5^{MHV}(34125)-M_5^{MHV}(41235)\cr
&=&V(x_1,x_2,(x_2-x_4+x_5),(x_3-x_4+x_5),x_5)\cr
&&-V((x_1-x_4+x_3),x_1,x_2,(x_2+x_5-x_4),(x_3+x_5-x_4))\cr
&&+V(x_3,(x_1+x_3-x_5),(x_1+x_4-x_5),x_1,x_2)\cr
&&-V((x_4-x_5+x_1),x_1,x_2,x_3,x_4)\cr
A_5^{MHV}(13245)&:&M_5^{MHV}(23145)-M_5^{MHV}(31245)+M_5^{MHV}(43125)-M_5^{MHV}(42315)\cr
&=&V(x_1,(x_1-x_2+x_3),(x_1-x_2+x_4),x_4,x_5)\cr
&&-V(x_1,(x_1-x_3+x_4),(x_2-x_3+x_4),x_4,x_5)\cr
&&+V(x_4,(x_1+x_4-x_5),x_1,(x_1-x_3+x_4),(x_2-x_3+x_4))\cr
&&-V((x_4-x_5+x_1),x_1,(x_1-x_2+x_3),(x_1-x_2+x_4),x_4)\cr
A_5^{MHV}(13425)&:&M_5^{MHV}(23145)-M_5^{MHV}(31425)+M_5^{MHV}(43125)-M_5^{MHV}(24315)\cr
&=&V(x_1,(x_1-x_2+x_3),(x_1-x_2+x_4),x_4,x_5)\cr
&&-V((x_5-x_2+x_3),(x_1-x_2+x_3),(x_1-x_2+x_4),x_4,x_5)\cr
&&+V(x_4,(x_1+x_4-x_5),x_1,(x_1-x_3+x_4),(x_2-x_3+x_4)\cr
&&-V(x_4,x_5,x_1,(x_1-x_3+x_4),(x_2-x_3+x_4)\cr
A_5^{MHV}(14325)&:&M_5^{MHV}(23145)-M_5^{MHV}(23415)+M_5^{MHV}(34125)-M_5^{MHV}(31425)\cr
&=&V(x_1,(x_1-x_2+x_3),(x_1-x_2+x_4),x_4,x_5)\cr
&&-V((x_1-x_2+x_3),(x_1-x_2+x_4),x_4,x_5,(x_3+x_5-x_2))\cr
&&+V((x_3+x_5-x_1),x_3,x_4,x_5,(x_2+x_5-x_1))\cr
&&-V(x_2,x_3,x_4,x_5,(x_5-x_1+x_2))
\eea

The points for each amplitude are: 8 points for $A_5^{MHV}(12345)$:
\be
x_1,x_2,x_3,x_4,x_5,(x_1+x_4-x_5),(x_2+x_4-x_3),(x_1+x_4-x_3)
\ee

9 points for $A_5^{MHV}(12435)$:
\be
x_1,x_2,x_3,x_4,x_5,(x_3+x_5-x_4),(x_1+x_3-x_4),(x_1+x_3-x_5),(x_1+x_4-x_5)
\ee
10 points for $A_5^{MHV}(14235)$:
\be
x_1,x_2,x_3,x_4,x_5,(x_2+x_5-x_4),(x_3+x_5-x_4),(x_1+x_3-x_4),(x_1+x_3-x_5),(x_1+x_4-x_5)
\ee
8 points for $A_5^{MHV}(13245)$:
\be
x_1,x_4,x_5,(x_1+x_3-x_2),(x_1+x_4-x_2),(x_1+x_4-x_3),(x_2+x_4-x_3),(x_1+x_4-x_5)
\ee
8 points for $A_5^{MHV}(13425)$:
\be
x_1,x_4,x_5,(x_1+x_3-x_2),(x_1+x_4-x_2),(x_1+x_4-x_3),(x_2+x_4-x_3),(x_1+x_4-x_5)
\ee
10 points for $A_5^{MHV}(14325)$:
\be
x_1,x_2,x_3,x_4,x_5,(x_1+x_3-x_2),(x_1+x_4-x_2),(x_2+x_5-x_1),(x_3+x_5-x_1),(x_3+x_5-x_2)
\ee

The expansion of the $6$-point tree amplitudes in the KK basis (KK relations) is explicitly
\bea
A_6(512346)&=&-A_6(123456)-A_6(123546)-A_6(125346)-A_6(152346)\cr
A_6(234156)&=&-A_6(143256)-A_6(143526)-A_6(145326)-A_6(154326)\cr
A_6(341256)&=&A_6(125436)+A_6(124536)+A_6(142536)\cr
&&+A_6(124356)+A_6(142356)+A_6(143256)\cr
A_6(341526)&=&A_6(152436)+A_6(154236)+A_6(145236)\cr
&&+A_6(154326)+A_6(145326)+A_6(143526)\cr
A_6(345126)&=&-A_6(154326)-A_6(154236)-A_6(152436)-A_6(125436)\cr
A_6(354126)&=&-A_6(145326)-A_6(145236)-A_6(142536)-A_6(124536)\cr
A_6(534126)&=&-A_6(143526)-A_6(143256)-A_6(142356)-A_6(124356)\cr
A_6(412356)&=&-A_6(123546)-A_6(123456)-A_6(124356)-A_6(142356)\cr
A_6(412536)&=&-A_6(125346)-A_6(125436)-A_6(124536)-A_6(142536)\cr
A_6(415236)&=&-A_6(152346)-A_6(152436)-A_6(154236)-A_6(145236)\cr
A_6(451236)&=&A_6(123546)+A_6(125346)+A_6(152346)\cr
&&+A_6(125436)+A_6(152436)+A_6(154236)\cr
A_6(541236)&=&A_6(123456)+A_6(124356)+A_6(142356)\cr
&&+A_6(124536)+A_6(142536)+A_6(145236)\cr
&&
\eea

\section{IR divergences of leading NMHV 6-point amplitudes}

In this Appendix we will review for completeness the IR divergences of leading NMHV 6-point amplitudes, in the massive regularization given in 
\cite{Mason:2010pg} for the MHV $n$-point case, and introduced in \cite{Alday:2009zm,Henn:2010bk,Henn:2010ir}.
 For the MHV case, only the $F_{1m}$ and $F_{2me}$ boxes were needed, and they were given in 
\cite{Mason:2010pg}. With the choice $X_i\cdot I=1, \forall i$, we have
\bea
F_{2me}(i-1,i,j-1,j)&=&-\log\Big(\frac{X_i\cdot X_j}{\mu^2}\Big)\log\Big(\frac{X_{i-1}\cdot X_{j-1}}{\mu^2}\Big)+\frac{1}{2}\log^2\Big(\frac{
X_{i}\cdot X_{j-1}}{\mu^2}\Big)\cr
&&+\frac{1}{2}\log^2\Big(\frac{X_{i-1}\cdot X_{j}}{\mu^2}\Big)+{\rm Li}_2\Big(1-\frac{X_i\cdot X_{j-1}}{X_{i-1}\cdot X_{j-1}}\Big)\cr
&&+{\rm Li}_2\Big(1-\frac{X_i\cdot X_{j-1}}{X_{i}\cdot X_{j}}\Big)
+{\rm Li}_2\Big(1-\frac{X_{i-1}\cdot X_{j}}{X_{i-1}\cdot X_{j-1}}\Big)\cr
&&+{\rm Li}_2\Big(1-\frac{X_{i-1}\cdot X_{j}}{X_{i}\cdot X_{j}}\Big)
+{\rm Li}_2\Big(1-\frac{X_i\cdot X_{j-1} X_{i-1}\cdot X_j}{X_i\cdot X_j X_{i-1}\cdot X_{j-1}}\Big)\cr
F_{1m}(i-3,i-2,i-1,i)&=&-\log\Big(\frac{X_{i-3}\cdot X_{i-1}}{\mu^2}\Big)\log\Big(\frac{X_{i-2}\cdot X_i}{\mu^2}\Big)
+\frac{1}{2}\log^2\Big(\frac{X_i\cdot X_{i-3}}{\mu^2}\Big)\cr
&&+{\rm Li}_2\Big(1-\frac{X_i\cdot X_{i-3}}{X_{i-3}\cdot X_{i-1}}\Big)+{\rm Li}_2\Big(1-\frac{X_i\cdot X_{i-3}}{X_{i-2}\cdot X_{i}}\Big)
+\frac{\pi^2}{6}\label{msfs}
\eea
We check that this gives the correct IR divergence for the 6-point MHV amplitude. We first extract the $\mu\rightarrow 0$ divergence of 
(\ref{msfs}) and obtain in this limit (using $\log(A/\mu^2)\log(B/\mu^2)=1/2 \log (A/\mu^2)
+1/2\log(B/\mu^2)+$finite)
\bea
F_{1m}(i-1,i,j-1,j)&=&\frac{1}{2}\log^2\Big(\frac{X_{i-3}\cdot X_{i-1}}{\mu^2}\Big)+\frac{1}{2}\log^2\Big(\frac{X_{i-2}\cdot X_i}{\mu^2}\Big)
\cr
&&+\frac{1}{2}\log^2\Big(\frac{X_i\cdot X_{i-3}}{\mu^2}\Big)+{\rm finite}\cr
F_{2me}(i-1,i,j-1,j)&=&\frac{1}{2}\log^2\Big(\frac{X_i\cdot X_{j-1}}{\mu^2}\Big)-\frac{1}{2}\log^2\Big(\frac{X_i\cdot X_{j}}{\mu^2}\Big)\cr
&&-\frac{1}{2}\log^2\Big(\frac{X_{i-1}\cdot X_{j-1}}{\mu^2}\Big)+{\rm finite}\cr
&&\label{massdiv}
\eea
Then for the MHV amplitude 
\bea
\frac{A_{6;1}^{MHV}(123456)}{A_{6;0}^{MHV}(123456)}&=&F_{2me}(1245)+F_{2me}(2356)+F_{2me}(3461)\cr
&&+F_{1m}(1234)+F_{1m}(2345)+F_{1m}(3456)\cr
&&+F_{1m}(4561)+F_{1m}(5612)+F_{1m}(6123)\cr
&&
\eea
we obtain the IR divergence in $\mu$ regularization
\bea
&&\Big[\frac{1}{2}\log^2\left(\frac{X_2\cdot X_4}{\mu^2}\right)+\frac{1}{2}\log^2\left(\frac{X_5\cdot X_1}{\mu^2}\right)\cr
&&-\frac{1}{2}\log^2\left(\frac{X_2\cdot X_5}{\mu^2}\right)-\frac{1}{2}\log^2\left(\frac{X_1\cdot X_4}{\mu^2}\right)+2\; {\rm permutations}\Big]\cr
&&+\frac{1}{2}\log^2 \left(\frac{X_1\cdot X_4}{\mu^2}\right)-\frac{1}{2}\log^2\left(\frac{X_1\cdot X_3}{\mu^2}\right)-\frac{1}{2}\log^2\left(
\frac{X_2\cdot X_4}{\mu^2}\right)+5\; {\rm permutations}\cr
&=&-\frac{1}{2}\sum_{i=1}^6\log^2\left(\frac{X_i\cdot X_{i+2}}{\mu^2}\right)
\eea
Then this indeed agrees with the result of the dimensional regularization IR divergence
\be
-\frac{1}{\epsilon^2}\sum_{i=1}^6\Big(\frac{X_i\cdot X_{i+2}}{\mu^2}\Big)^{-\epsilon}
\ee
by dropping the terms divergent in $\epsilon$ and keeping the terms finite in $\epsilon$, but still divergent in $\mu\rightarrow 0$. 
Since the twistor space $X_i$ is only defined in 4 dimensions, whereas the rest of the 
formulas are in $d$ dimensions, the result is only formal. 

All the box functions in dimensional regularization on the dual space ($x_i$) were given in Appendix B of \cite{Elvang:2009ya}
(see also \cite{Drummond:2008bq}). 
We can translate to our notation and use the 4d twistor variables relation $X_i\cdot X_j=-(x_i-x_j)^2\equiv -x_{ij}^2$, obtaining in a formal way
\bea
F_{2me}(i-1,i,j-1,j)&=&-\frac{1}{\epsilon^2}\Big[\left(\frac{X_{i-1}\cdot X_{j-1}}{\mu^2}\right)^{-\epsilon}+\left(\frac{
X_i\cdot X_j}{\mu^2}\right)^{-\epsilon}-\left(\frac{X_i\cdot X_{j-1}}{\mu^2}\right)^{-\epsilon}\cr
&&-\left(\frac{X_{i-1}\cdot X_j}{\mu^2}\right)^{-\epsilon}\Big]
+\frac{1}{2}\log^2\Big(\frac{X_{i-1}\cdot X_{j-1}}{X_i\cdot X_j}\Big)\cr
&&+{\rm Li}_2\Big(1-\frac{X_i\cdot X_{j-1}}{X_{i-1}\cdot X_{j-1}}\Big)
+{\rm Li}_2\Big(1-\frac{X_i\cdot X_{j-1}}{X_{i}\cdot X_{j}}\Big)\cr
&&+{\rm Li}_2\Big(1-\frac{X_{i-1}\cdot X_{j}}{X_{i-1}\cdot X_{j-1}}\Big)+{\rm Li}_2\Big(1-\frac{X_{i-1}\cdot X_{j}}{X_{i}\cdot X_{j}}\Big)\cr
&&+{\rm Li}_2\Big(1-\frac{X_i\cdot X_{j-1} X_{i-1}\cdot X_j}{X_i\cdot X_j X_{i-1}\cdot X_{j-1}}\Big)+{\cal O}(\epsilon)\cr
F_{1m}(i-3,i-2,i-1,i)&=&-\frac{1}{\epsilon^2}\Big[\left(\frac{X_{i-1}\cdot X_{i-3}}{\mu^2}\right)^{-\epsilon}+\left(\frac{X_{i-2}\cdot 
X_i}{\mu^2}\right)^{-\epsilon}-\left(\frac{X_i\cdot X_{i-3}}{\mu^2}\right)^{-\epsilon}\Big]\cr
&&+\frac{1}{2}\log^2\Big(\frac{X_{i-1}\cdot X_{i-3}}{X_{i-2}\cdot X_i}\Big)+{\rm Li}_2\Big(1-\frac{X_i\cdot X_{i-3}}{X_{i-3}\cdot X_{i-1}}\Big)\cr
&&+{\rm Li}_2\Big(1-\frac{X_i\cdot X_{i-3}}{X_{i-2}\cdot X_{i}}\Big)+\frac{\pi^2}{6}+{\cal O}(\epsilon)\cr
F_{2mh}(i-1,i,i+1,j)&=&-\frac{1}{2\epsilon^2}\Big[\left(\frac{X_{i-1}\cdot X_{i+1}}{\mu^2}\right)^{-\epsilon}+2\left(\frac{X_i\cdot X_j}{\mu^2}\right)
^{-\epsilon}\cr
&&-\left(\frac{X_{i+1}\cdot X_j}{\mu^2}\right)^{-\epsilon}-\left(\frac{X_{i-1}\cdot X_j}{\mu^2}\right)^{-\epsilon}\Big]\cr
&&+{\rm Li}_2\Big(1-\frac{X_{i+1}\cdot X_j}{X_i\cdot X_j}\Big)+{\rm Li}_2\Big(1-\frac{X_{i-1}\cdot X_j}{X_i\cdot X_j}\Big)\cr
&&-\frac{1}{2}\log\Big(\frac{X_{i+1}\cdot X_j}{X_{i-1}\cdot X_{i+1}}\Big)\log\Big(\frac{X_{i-1}\cdot X_j}{X_{i-1}\cdot X_i+1}\Big)\cr
&&+\frac{1}{2}\log^2\Big(\frac{X_{i-1}\cdot X_{i+1}}{X_i\cdot X_j}\Big)+{\cal O}(\epsilon)\cr
&&
\eea
We can then check immediately that by expanding $F_{1m}$ and $F_{2me}$ in $\epsilon$ and keeping only the terms finite in $\epsilon$, 
but still divergent in $\mu\rightarrow 0$, we get the same result as in (\ref{msfs}), confirming our formal manipulations. 
Then we do the same for $F_{2mh}$, and obtain 
\bea
F_{2mh}(i-1,i,i+1,j)&=&\frac{1}{4}\Big[\log^2\left(\frac{X_{i-1}\cdot X_{i+1}}{\mu^2}\right)+\log^2\left(\frac{X_{i+1}\cdot X_j}
{\mu^2}\right)+\log^2\left(\frac{X_{i-1}\cdot X_j}{\mu^2}\right)\Big]\cr
&&-\log\left(\frac{X_{i-1}\cdot X_{i+1}}{\mu^2}\right)\log \left(\frac{X_i\cdot X_j}{\mu^2}\right)\cr
&&+{\rm Li}_2\Big(1-\frac{X_{i+1}\cdot X_j}{X_i\cdot X_j}\Big)+{\rm Li}_2\Big(1-\frac{X_{i-1}\cdot X_j}{X_i\cdot X_j}\Big)
\eea
which is correct, and is the 4 dimensional $F_{2mh}$ in $\mu$ regularization. 
Then the $\mu\rightarrow 0$ IR divergence of the $F_{1m}(1234)+F_{1m}(4561)$ terms in $W_6^{(1)}$ is from (\ref{massdiv}) 
\bea
&&-\frac{1}{2}\log^2\left(\frac{X_1\cdot X_4}{\mu^2}\right)-\frac{1}{2}\log^2\left(\frac{X_1\cdot X_3}{\mu^2}\right)+\frac{1}{2}\log^2\left(
\frac{X_2\cdot X_4}{\mu^2}\right)\cr
&&-\frac{1}{2}\log^2\left(\frac{X_1\cdot X_4}{\mu^2}\right)-\frac{1}{2}\log^2\left(\frac{X_4\cdot X_6}{\mu^2}\right)+\frac{1}{2}\log^2\left(
\frac{X_5\cdot X_1}{\mu^2}\right)
\eea
and the IR divergence of the $F_{2mh}(6124)$ and $F_{2mh}(3451)$ terms is
\bea
&&\frac{1}{4}\Big(\log^2\left(\frac{X_2\cdot X_4}{\mu^2}\right)+\log^2\left(\frac{X_4\cdot X_6}{\mu^2}\right)-\log^2\left(\frac{
X_2\cdot X_6}{\mu^2}\right)\Big)-\frac{1}{2}\log^2\left(\frac{X_1\cdot X_4}{\mu^2}\right)\cr
&&\frac{1}{4}\Big(\log^2\left(\frac{X_5\cdot X_1}{\mu^2}\right)+\log^2\left(\frac{X_1\cdot X_3}{\mu^2}\right)-\log^2\left(\frac{
X_5\cdot X_3}{\mu^2}\right)\Big)-\frac{1}{2}\log^2\left(\frac{X_1\cdot X_4}{\mu^2}\right)
\eea
Adding them up we get the IR divergence of $W_6^{(1)}$,
\be
-\frac{1}{4}\sum_{i=1}^6\log^2\left(\frac{X_i\cdot X_{i+2}}{\mu^2}\right)
\ee
Since this result is cyclically invariant, we get the same for $W_6^{(2)}$ and $W_6^{(3)}$. The IR divergence of the NMHV 6-point leading 
amplitude in $\mu$ regularization is then 
\be
\frac{1}{2}(B_1+B_2+B_3)W_{1,div}=A_{tree}^{NMHV}W_{1,div}=-\frac{A_{tree}^{NMHV}}{4}\sum_{i=1}^6\log^2\left(\frac{X_i\cdot X_{i+2}}{\mu^2}\right)
\ee
which is indeed compatible with the known result in dimensional regularization 
\be
-\frac{1}{2\epsilon^2}A_{tree}^{NMHV}\sum_{i=1}^6\left(-\frac{x_{i,i+2}^2}{\mu^2}\right)^{-\epsilon}
\ee

\end{appendix}
\newpage

\bibliographystyle{utphys}
\bibliography{twistorpaper}

\end{document}